\documentclass[final]{pasj00}
\usepackage{graphicx}

\SetRunningHead
{T. Akahori and K. Yoshikawa}
{Hydrodynamic Simulations of Merging Galaxy Clusters}

\title{Hydrodynamic Simulations of Merging Galaxy Clusters: 
Non-Equilibrium Ionization State and Two-Temperature Structure}

\author{Takuya \textsc{Akahori}$^1$ and Kohji \textsc{Yoshikawa}$^2$}
\affil{$^1$Research Institute of Basic Science, Chungnam National University,
Daejeon 305-764, Korea}
\email{akataku@canopus.cnu.ac.kr}
\affil{$^2$Center for Computational Sciences, University of Tsukuba, 1-1-1, 
Tennodai, Tsukuba, Ibaraki 305-8577}
\email{kohji@ccs.tsukuba.ac.jp}


\Received{2009 September 21}
\Accepted{2010 January 20}
\Published{}

\begin{document}
\maketitle
\begin{abstract}
 We investigate a non-equilibrium ionization state and an electron-ion
 two-temperature structure of the intracluster medium (ICM) in
 merging galaxy clusters using a series of N-body and hydrodynamic
 simulations. Mergers with various sets of mass ratios and impact
 parameters are systematically investigated, and it is found that, in
 most cases, ICM significantly departs from the
 ionization equilibrium state at the shock layers with a Mach number
 of $\sim$1.5--2.0 in the outskirts of the clusters, and the shock
 layers with a Mach number of $\sim$2--4 in front of the ICM cores.
 Accordingly, the intensity ratio between
 Fe\,\textsc{xxv} and Fe\,\textsc{xxvi} K$\alpha$ line emissions is
 significantly altered from that in the ionization equilibrium state.
 If the effect of the two-temperature structure of ICM is
 incorporated, the electron temperature is $\sim$10--20~\% and
 $\sim$30--50~\% lower than the mean temperature of ICM at the
 shock layers in the outskirts and in front of the ICM cores,
 respectively, and the deviation from the ionization equilibrium
 state becomes larger. We also address the dependence of the
 intensity ratio on the viewing angle with respect to the merging
 plane.
\end{abstract}

\section{Introduction}
Galaxy clusters host an X-ray emitting hot gas with temperature of
$\sim 10^{7-8}$ K, or the intracluster medium (ICM), as well as dark
matter and galaxies. It is theoretically believed that ICM has been
heated up through various magnitude of shock waves caused by
successive mergers of galaxies, galaxy groups and galaxy clusters
during the hierarchical structure formation (e.g., \cite{Ryu03}).
These shocks are regarded as sources of turbulent motion of ICM and
sites of particle acceleration.  Furthermore, they amplify the
intracluster magnetic field and can be sources of non-thermal
emission.  In this sense, merging galaxy clusters are of great
importance as fruitful sites of interesting phenomena of astrophysical
plasma (see \cite{Sarazin02} for a review).

From an observational point of view, merging galaxy clusters provide lots
of information about outskirts of galaxy clusters, because compression
and shock heating of ICM during merging events increase the X-ray
luminosity of ICM at the outer parts of galaxy clusters.  For example,
based on the Suzaku XIS observation, \citet{Fujita08} reported the
detection of Fe K emission lines in the linked region of Abell 399 and
Abell 401, more than $\sim 1$~Mpc apart from the centers of both
galaxy clusters, and estimated the metallicity relatively higher than
conventionally expected. Their detection is in virtue of the fact that
ICM in the linked region is significantly compressed, as well as the
high sensitivity of the XIS detector.  Therefore, merging galaxy
clusters can be good sites for elucidating physical properties of ICM,
such as the temperature and metallicity, especially in the peripheral
regions of galaxy clusters.

Measurements of physical properties of ICM, such as the temperature and
metallicity, in X-ray observations, however, usually assume that ICM
is in the ionization equilibrium state, and that electrons and ions
share the same thermal temperature. It is widely believed that these
assumptions are reasonable inside galaxy clusters, because the
timescale for collisional ionization equilibrium,
\begin{equation}\label{tcie}
t_{\rm CIE}\sim 
3\times 10^{7}~{\rm yr}\left(\frac{n_{\rm e}}{10^{-3}~{\rm cm^{-3}}}\right)^{-1},
\end{equation}
(for e.g., Fe\,\textsc{xxv} and Fe\,\textsc{xxvi} in a temperature of several keV; see also \cite{Masai84}) and the timescale for thermal equilibration
between electrons and ions through Coulomb scattering,
\begin{equation}\label{tei}
t_{\rm ei}\sim
7\times 10^{7}~{\rm yr}\left(\frac{n_{\rm e}}{10^{-3}~{\rm cm^{-3}}}
\right)^{-1}\left(\frac{T_{\rm e}}{5\times10^7~{\rm K}}\right)^{3/2},
\end{equation}
(\cite{Spitzer62}) are usually shorter than the typical dynamical
timescale around the central regions of galaxy clusters, where $n_{\rm
e}$ and $T_{\rm e}$ are the number density and temperature of
electrons, respectively. But these timescales are not short enough
compared with the merger timescale, and even longer in the outskirts
of galaxy clusters where $n_{\rm e}\sim 10^{-4}$\,cm$^{-3}$.
Consequently, the assumptions of the ionization equilibrium and the
thermal equipartition between electrons and ions are no longer
justified.  Therefore, in measuring the temperature and/or metallicity
of ICM around shock layers and outskirts of merging galaxy clusters,
the deviations from the ionization equilibrium and the thermal
equipartition have to be taken into consideration 
(see also \cite{Prokhorov09}).

Previously, many works were devoted to numerical simulations of
merging galaxy clusters under idealized but well-controlled initial
conditions \citep{Roettiger97, Ricker01, Ritchie02, Takizawa05,
 Takizawa06}. But most of them ignored the effects of the
non-equilibrium ionization state and the deviation from thermal
equipartition between electrons and ions, or the two-temperature
structure of ICM. The former effect in the Warm--Hot Intergalactic 
Medium (WHIM) is addressed in
cosmological simulations for the first time by \citet{YS06} and
\citet{CF06}. \citet{Takizawa99} and \citet{Takizawa00} took into
account the latter effect in numerical simulations of merging galaxy
clusters for the first time, while \citet{Yoshida05} and \citet{RN09}
investigated it in WHIM and in
outskirts of galaxy clusters, respectively, in cosmological
simulations.  Recently, \citet{AY08}(AY08) carried out an N-body/SPH
simulation of an Abell 399 and Abell 401 system incorporating these two
effects simultaneously and self-consistently, and showed that the
non-equilibrium ionization state and the two-temperature structure of
ICM take place along shock layers at the merging interface.

While AY08 addressed only a particular situation of Abell 399 and
Abell 401, in this paper, we systematically investigate the
non-equilibrium ionization state and the two-temperature structure of
ICM for various situations of merging galaxy clusters, e.g., mass
ratios, impact parameters, and merging stages, so as to be confronted
with future X-ray observations of actual merging galaxy clusters. In
addition to that, we also address the viewing angle dependence of
observational signatures of the deviations from the ionization
equilibrium and the thermal equipartition.

The rest of this paper is organized as follows. In section 2, we
describe the numerical methods and the initial conditions of the
simulations. The simulation results are presented in section
3.  In section 4, we summarize our conclusions and discuss future
prospects for the detection of the non-equilibrium ionization state and
the two-temperature structure of ICM. In the case that the
cosmological scaling is required, we assume the density parameter
$\Omega_{\rm M}=0.24$, the baryon density parameter $\Omega_{\rm
b}=0.04$, and the Hubble constant $H_0 = 70$ km s$^{-1}$ Mpc$^{-1}$,
unless otherwise specified.

\section{Simulation}

\subsection{Numerical Method}
We carry out a set of N-body and SPH simulations of merging galaxy
clusters for various initial conditions. In all simulations, radiative
cooling is ignored, since, in this work, we focus on relatively outer
regions rather than the very central regions of galaxy clusters, and
the actual cooling timescale in the central regions of the simulated
ICM is $\sim 5$~Gyr, sufficiently longer than the merging timescale.
Simulations are performed using the code developed in the previous
papers (\cite{AM05}, \cite{AM06}, and AY08), which adopts the entropy
conservative formulation of SPH by \citet{SH02}, and the standard
Monaghan-Gingold artificial viscosity \citep{MG83} with Balsara
limiter (\cite{Balsara95}). In addition to the standard SPH
prescription, the time evolution of the two-temperature structure and
the non-equilibrium ionization state of ICM are incorporated.

For the time evolution of the two-temperature structure, only the
Coulomb scattering is considered as a physical process responsible for
the thermal relaxation between electrons and ions. In this picture,
timescales on which each of electrons and ions achieves the thermal
relaxation are much shorter than that between electrons and ions.
Therefore, electrons and ions can have different temperatures, $T_{\rm
e}$ and $T_{\rm i}$, respectively, just after experiencing shock
heating, because nearly all of the kinetic energy of ICM is carried by
ions at a shock front, and is preferentially converted to the thermal
energy of ions in the post-shock regions.  Afterward, thermal
relaxation between electrons and ions proceeds through Coulomb
scattering on a timescale described by
\begin{eqnarray}\label{tei2}
t_{\rm ei} 
&=&\frac{3m_{\rm e}m_{\rm i}}
{8(2\pi)^{1/2}n_{\rm i} Z_{\rm i}^2e^4 \ln \Lambda}
\left(\frac{k_{\rm B}T_{\rm e}}{m_{\rm e}}+\frac{k_{\rm B}T_{\rm i}}{m_{\rm i}}
\right)^{3/2} \\\nonumber
&\simeq& 2\times 10^8~{\rm yr}
\frac{(T_{\rm e}/10^8~{\rm K})^{3/2}}{(n_{\rm i}/10^{-3}~{\rm cm^{-3}})}
\cdot\left(\frac{40}{\ln\Lambda}\right),
\end{eqnarray}
where $m_{\rm e}$ and $m_{\rm i}$ are the electron and ion masses,
respectively, $n_{\rm i}$ the number density of ions, $k_{\rm B}$ the
Boltzmann constant, and $\ln\Lambda$ the Coulomb logarithm
\citep{Spitzer62}.

Assuming that electrons and ions share the same fluid velocity, and
that shock heating are exclusively contributed to the heating of ions,
the energy equations for electrons and ions are given as
\begin{equation}\label{eq2energy_e}
\rho\frac{d u_{\rm e}}{d t}=-P_{\rm e}\nabla 
\cdot v + Q_{\rm ex},
\end{equation}
\begin{equation}\label{eq2energy_i}
\rho\frac{d u_{\rm i}}{d t}=-P_{\rm i}\nabla 
\cdot v + Q_{\rm sh} - Q_{\rm ex},
\end{equation}
respectively, where $\rho$ is the gas mass density, $u$ the specific
thermal energy, $P$ the thermal pressure (quantities with subscript
``e'' and ``i'' denote those of electrons and ions, respectively),
$Q_{\rm sh}$ the shock heating rate per unit volume, and $Q_{\rm ex}$
the energy exchange rate per unit volume through Coulomb scattering:
\begin{equation}\label{Q_ex}
Q_{\rm ex}=\frac{n_{\rm e}k}{\gamma-1}
\frac{T_{\rm i}-T_{\rm e}}{t_{\rm ei}},
\end{equation}
where $\gamma=5/3$ is the adiabatic index. Using equations
(\ref{eq2energy_e}) and (\ref{eq2energy_i}), and assuming the constant
mean molecular weight, $\mu=0.6$, for almost fully ionized ICM, we
obtain an ordinary energy equation for mixed fluid with a mean
temperature, $\bar{T}=(n_{\rm e}T_{\rm e}+n_{\rm i}T_{\rm i})/(n_{\rm
e} +n_{\rm i})$, and another equation for the time evolution of
electron temperature, which can be written as
\begin{equation}\label{eq:eq2T}
\frac{d\tilde{T}_{\rm e}}{dt}=
\frac{\tilde{T}_{\rm i}-\tilde{T}_{\rm e}}{t_{\rm ei}}
-\frac{\tilde{T}_{\rm e}}{u}Q_{\rm sh},
\end{equation}
where $\tilde{T}_{\rm e}\equiv T_{\rm e}/\bar{T}$ and $\tilde{T}_{\rm
i}\equiv T_{\rm i}/\bar{T}$ are the dimensionless temperatures of
electrons and ions normalized by the mean temperature, respectively,
and $u=u_{\rm e}+u_{\rm i}$.  This equation is numerically solved in
the same manner described in \citet{Takizawa99} as follows. We first
regard the second term on the right hand side as being negligibly
small, and integrate the first term implicitly using the method
described in \citet{FL97}. Then we integrate the second term
explicitly, and finally add this contribution to the time evolution of
$\tilde{T}_{\rm e}$.  In each integration, the time step regulated by
the courant condition of SPH ensures sufficiently smaller $Q_{\rm
sh}/u$ than unity and thus enable us to solve the evolution of
$\tilde{T}_{\rm e}$ with reasonable accuracy.

Actually, a possibility is argued that plasma waves are able to attain
the thermal relaxation between electrons and ions more quickly than
the Coulomb collision. In reality, it is observationally suggested
that the electron and ion temperatures are almost the same around the
shock front with a Mach number of $\simeq 10-20$ in supernova
remnants, indicating that physical processes other than Coulomb
collision contribute to the quick thermal relaxation between electrons
and ions \citep{Ghavamian07}. There exists, however, a room for a very
slow shock with a Mach number of $M\lesssim 5$ to have significant
difference in the temperatures between electrons and ions. Thus, in
order to bracket the plausible range of theoretical uncertainties, we
present the results both for the two-temperature runs described above
and the single-temperature runs, in which thermal energies of electrons 
and ions are assumed to be instantaneously equilibrated, and electrons 
and ions share a single temperature.

The time evolution of ionization fractions of ions is computed by
solving the rate equation,
\begin{eqnarray}\label{eq:noneq}
\frac{df_j}{dt}&=&\sum_{k=1}^{j-1}S_{j-k,k}f_k-
\sum_{i=j+1}^{Z+1}S_{i-j,j}f_j\nonumber\\
&-&\alpha_jf_j+\alpha_{j+1}f_{j+1},
\end{eqnarray}
for each SPH particle, where $j$ is the index of a particular
ionization stage considered, $Z$ the atomic number, $f_j$ the
ionization fraction of an ion $j$, $S_{i,j}$ the ionization rate of an
ion $j$ with the ejection of $i$ electrons, and $\alpha_j$ is the
recombination rate of an ion $j$.  The ionization processes include
collisional, Auger, charge-transfer, and photo-ionizations, and
recombination processes are composed of radiative and dielectronic
recombinations.  The ionization and recombination rates are calculated
by utilizing the SPEX ver 1.10 software
package\footnote{http://www.sron.nl/divisions/hea/spex/}.

Equation~(\ref{eq:noneq}) is integrated implicitly in the same manner
described in \citet{YS06}, in which ionization fractions of all ionic
species are sequentially updated in the order of increasing ionization
states rather than being updated simultaneously, and the source terms
contributed by the ionization from and recombination to the lower
states are evaluated at the advanced time step.  In the
two-temperature runs, the reaction rates are computed using the
electron temperature, $T_{\rm e}$, calculated from
equation~(\ref{eq:eq2T}) so as to incorporate the effect of the
two-temperature structure.  As for the single-temperature runs, the
reaction rates are computed using the mean temperature, $\bar{T}$.

We solve the time evolution of the ionization fractions of H, He, C,
N, O, Ne, Mg, Si, S, and Fe; total 112 ionization states for each SPH
particle.  All the ionization states are considered in calculations of
the X-ray spectrum and the surface brightness in this paper. But in
the following, we mainly focus on the ionization fraction and the line
intensities of Fe, because the line emissions of Fe are the most
prominent in the typical temperature range of ICM $\sim 10^{7-8}$~K.

\subsection{Initial Condition}
The initial conditions are setup as follows. We adopt the NFW density
profile \citep{NFW97} for a dark matter halo, and the $\beta$-model
density profile \citep{CF76} with $\beta = 2/3$ for a ICM component.
The scale radius of the NFW profile is set to $r_{\rm s} = r_{\rm
200}/5.16$, and the core radius of the $\beta$-model is $r_{\rm
c}=r_{\rm s}/2$, where $r_{\rm 200}$ is the virial radius defined as
the radius within which the mean cluster mass density is 200 times the
present cosmic critical density. The radial profiles of velocity
dispersion of dark matter and temperature of ICM are computed using
Jeans equation and the assumption of hydrostatic equilibrium,
respectively.  The gas and dark matter particles are distributed out
to $1.4~r_{\rm 200}$, which enables us to follow the dynamical
evolution of ICM at outskirts of galaxy clusters.

\begin{figure}[tbp]
\begin{center}
\FigureFile(70mm,35mm){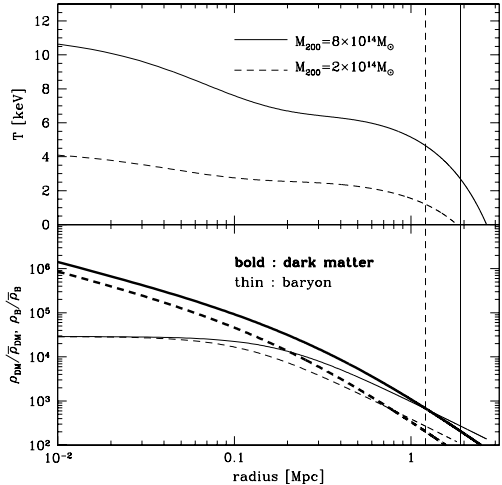}
\end{center}
\caption{\label{fig:profiles} Radial density profiles of ICM (baryon)
and dark matter normalized by the cosmic mean density of each
component (lower panel), and ICM temperature (upper panel) at
initial conditions. Solid lines indicate the profiles of galaxy
clusters in the runs with 1:1 mass ratio and the main cluster in the
runs with 4:1 mass ratio ($M_{200}=8\times 10^{14}\MO$), while
dashed lines show the profiles of the sub-cluster in the runs with
4:1 mass ratio ($M_{200}=2 \times 10^{14}\MO$). The vertical lines
indicate the locations of their virial radii. The central ICM
density corresponds to $n_{\rm e}=5.52\times 10^{-3}~{\rm
 cm^{-3}}$.\label{fig:profile}}
\end{figure}

In this paper, mass ratios of two galaxy clusters are set to 1:1 and
4:1. For runs with 1:1 mass ratio, the virial mass and radius are set
to $M_{\rm 200} = 8 \times 10^{14}~\MO$ and $r_{\rm 200}=1.91$~Mpc,
respectively. For runs with 4:1 mass ratio, a main (or more massive)
cluster is the same as the one in the runs with 1:1 mass ratio, and
the virial mass and radius of a sub (or less massive) cluster are set
to $M_{\rm 200} = 2 \times10^{14}~\MO$ and $r_{\rm 200}=1.21$~Mpc,
respectively. Density profiles of dark matter and ICM and temperature
profiles for the two masses of galaxy clusters are depicted in
figure~\ref{fig:profiles}.  For each mass ratio, we carry out the runs
with different impact parameters, $b$; ``head-on'' runs with $b=0$ and
``off-set'' runs with $b=0.955\,\mbox{Mpc}$.  As for the metallicity, a
spatially uniform metallicity of 0.2 times the solar abundance is
assumed. In addition to that, we also assume that electrons and ions
share the same temperature, and that they are in the ionization
equilibrium state at the start of the simulations.

\begin{table}
\caption{Simulation runs and their merging parameters\label{tab:runs}}
\begin{center}
\begin{tabular}{ccccc}
 \hline\hline
 \noalign{\smallskip}
 run ID & $M_1$ [$\MO$] & $M_2$ [$\MO$]& $v_{\rm i}$ [km s$^{-1}$] & $b$ [Mpc]\\
 \noalign{\smallskip}
 \hline
 \noalign{\smallskip}
 1:1 head on & $8\times 10^{14}$ & $8\times 10^{14}$ & 667 & 0.0 \\
 \noalign{\smallskip}
 1:1 offset  & $8\times 10^{14}$ & $8\times 10^{14}$ & 675 & 0.955 \\
 \noalign{\smallskip}
 4:1 head on & $8\times 10^{14}$ & $2 \times 10^{14}$ & 645 & 0.0 \\
 \noalign{\smallskip}
 4:1 offset  & $8\times 10^{14}$ & $2\times 10^{14}$ & 654 & 0.955 \\
 \noalign{\smallskip}
 \hline
\end{tabular}
\end{center}
\end{table}

The initial separation of two galaxy clusters, $d$, is set to 1.4
times of the sum of their virial radii so that they contact each other
at their outer edge. The initial relative velocity of the two
clusters, $v_{\rm i}$, is calculated using the prescription described
in \citet{Sarazin02}, and is given by
\begin{equation}
v_{\rm i}^2=2G(M_1+M_2)\left(\frac{1}{d}-\frac{1}{d_0}\right)\left[1-\left(\frac{b}{d_0}\right)^2\right]^{-1},
\end{equation}
where $G$ is the gravitational constant, and $M_1$ and $M_2$ are the
virial masses of the two clusters.  $d_0$ denotes the maximum
separation where their relative {\it radial} velocity is zero, and is
calculated with the aid of the Kepler's third law as follow
\begin{equation}
d_0 = [2G(M_1+M_2)]^{1/3}\left(\frac{t_{\rm 0}}{\pi}\right)^{2/3},
\end{equation}
where $t_0$ is the comic time when the merging event takes place, and
is set to $t_0 = 13.7$~Gyr in this paper.  All the runs presented in
this paper and their merging parameters are listed in
table~\ref{tab:runs}.

In all simulations, the mass of each dark matter and SPH particle is
set to $8.62 \times 10^8~M_{\odot}$ and $1.64 \times 10^8~M_{\odot}$,
respectively. The single cluster in the runs with 1:1 mass ratio, and
the main cluster in the runs with 4:1 mass ratio are composed of one
million dark matter particles and the same number of SPH particles,
while the sub-clusters in the runs with 4:1 mass ratio are represented
by a quarter million particles for each component. The corresponding 
spatial resolution (the smoothing length) of SPH is $\sim 50$~kpc at the 
ICM density of $10^{-3} {\rm cm^{-3}}$ (central part of the clusters), 
or $\sim 110$~kpc at $10^{-4} {\rm cm^{-3}}$ (outskirts of the clusters).

\section{Result}

\subsection{Single-Temperature Runs}\label{subsec1}

\begin{figure*}[htbp]
\begin{center}
 \FigureFile(160mm,80mm){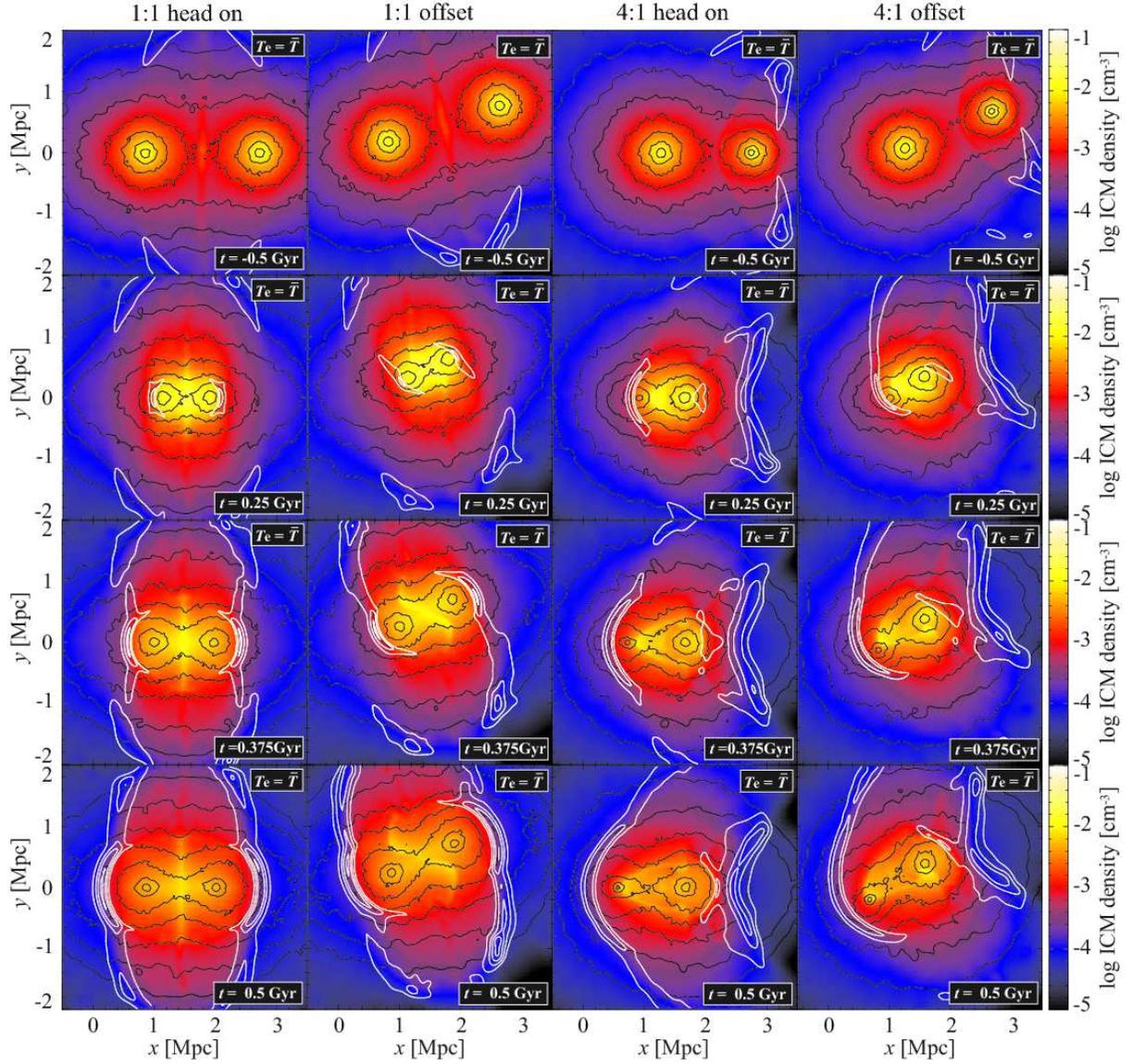}
\end{center}
\caption{Density maps of ICM on a collision plane of the two galaxy
 clusters are shown for 1:1 head on, 1:1 offset, 4:1 head on, and
 4:1 offset (from the left to the right) at a time of $t=-0.5$~Gyr,
 0.25~Gyr, 0.375~Gyr, and 0.5~Gyr (from the top to the bottom),
 respectively, where $t=0~\mbox{Gyr}$ corresponds to the time of the
 closest approach of the centers of the dark matter halos. White
 contours indicate the Mach number distribution of ICM for a Mach
 number of 1.4 and above, and are separated by a difference of 0.6.
 The innermost contours for 1:1 and 4:1 mergers at a time of 
 $t=0.5$~Gyr are 3.2 and 2.6, respectively. 
 The distributions of dark matter are also overlaid with black
 contours.\label{fig:result1_1T}}
\end{figure*}

First, we show the results for the single-temperature runs, in which
thermal relaxation between electrons and ions is supposed to be
achieved very quickly, and electrons and ions share a single
temperature.

Figure~\ref{fig:result1_1T} shows the time evolution of the ICM
density (color), the dark matter density (black contour), and the Mach
number distribution of ICM (white contour), on a collision plane of
the two merging galaxy clusters, where the Mach number is estimated
with the shock-capture scheme described in \citet{Pfrommer06}. Panels
from top to bottom show the snapshots at a time of $t=-0.5$~Gyr,
0.25~Gyr, 0.375~Gyr and 0.5~Gyr, respectively, where $t=0$~Gyr
corresponds to the time of the closest approach of the centers of the
dark matter halos. Therefore, dark matter halos at $t>0$ have already
passed each other.

In the early stage of the merging ($t=-0.5$~Gyr), the outskirts of the
two galaxy clusters interact with each other, and then form the linked
regions where ICM is compressed due to the collision. At the edges of
the linked regions, there exist shock layers with a Mach number of
1.5--2.0 (white contour in the top panels of
figure~\ref{fig:result1_1T}). These shocks are immediately generated
at the contact interface of the two clusters when the clusters start
to collide, then propagate to outside of each cluster.

The emergence of these shocks at the outskirts of galaxy clusters is
also pointed out by AY08, but not clearly seen in the previous SPH
simulations (\cite{Takizawa99}; \cite{Takizawa00}). This is because,
in these previous studies, density profiles of ICM are truncated in
the outside of clusters ($r>5r_{\rm c}$), and also because these
studies adopt isothermal temperature profiles of ICM at the start of
the simulations. On the contrary, in the present study, the
distributions of ICM are extended out to $r=14.4~r_{\rm c}$, and the
ICM temperature decreases toward outside as suggested by recent X-ray
observations (e.g., \cite{Vikhlinin05}). Therefore, the sound speed at
the outskirts of galaxy clusters, which is typically $\sim 500~\mbox{
km s$^{-1}$}$ for a gas temperature of $\sim 2~{\rm keV}$, is lower
than the relative velocity ($\sim$700--800~$\mbox{km s$^{-1}$}$) 
of the two clusters at this stage, resulting in the formation of the shock
waves.

At the middle stage of the merging ($t=0.25$--$0.375~{\rm Gyr}$), 
another shock waves with a Mach number of 1.5--2.0 in front of 
the ICM density peaks emerge, because the dense cores of ICM are 
accelerated by the preceding dark matter halos, and they have Mach 
numbers of $\sim$3--4 and 2--3 at the late stage of the merging 
($t=0.5~{\rm Gyr}$) in the case of 1:1 and 4:1 mass ratios, respectively.
Since the shock waves formed in front of the ICM cores are
faster than the ones formed in the outskirts at the early stage, the 
former overtake and merge with the latter, as can be seen in the bottom 
panels of figure~\ref{fig:result1_1T}.

\begin{figure*}[htbp]
\begin{center}
 \FigureFile(160mm,80mm){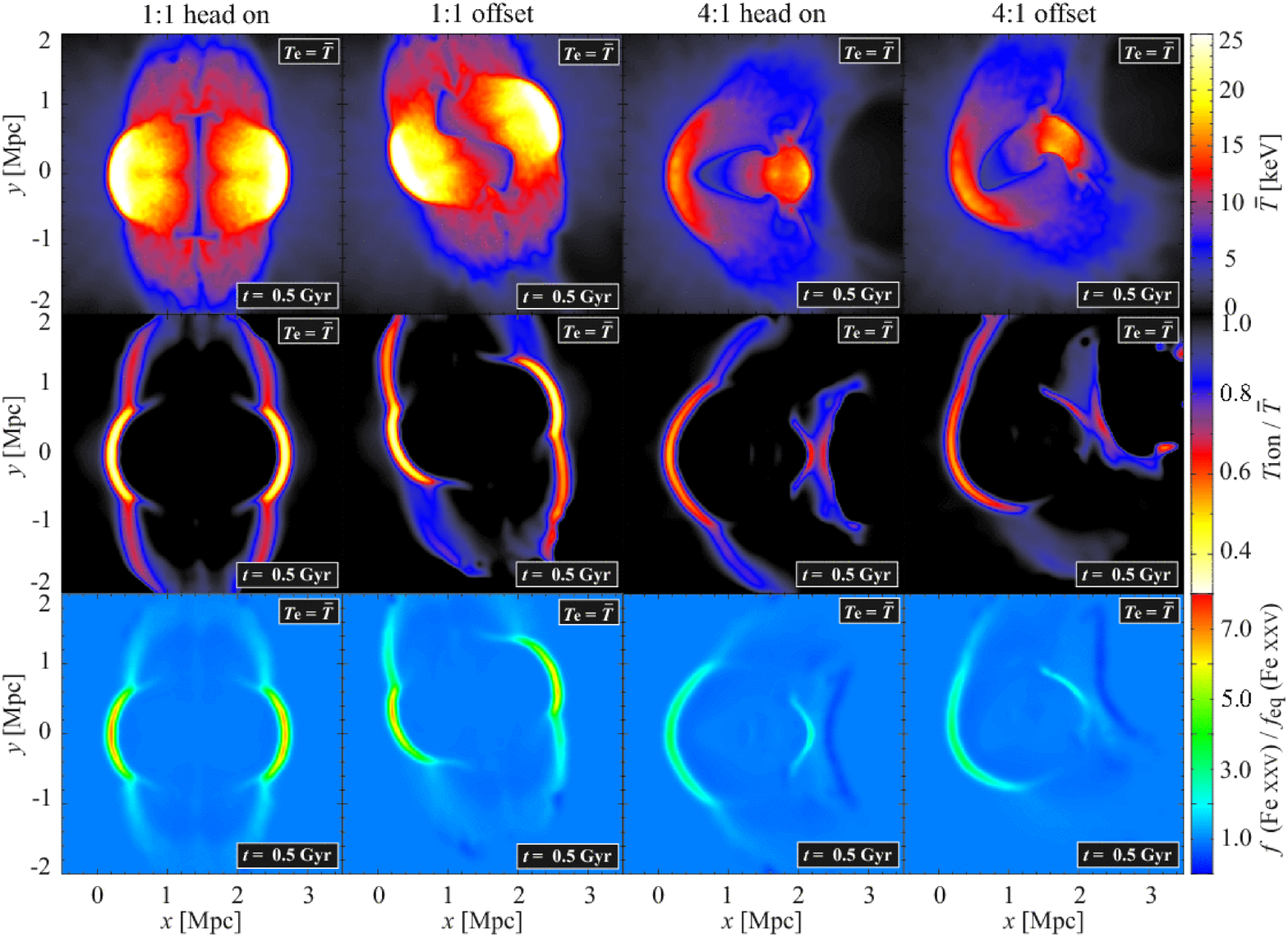}
\end{center}
\caption{The mean temperature of ICM, $\bar{T}$, the ratio of the
 ionization temperature relative to the mean temperature,
 $T_{\rm ion}/\bar{T}$, and the ratio of the ionization fraction of
 Fe\,\textsc{xxv} relative to that in the ionization equilibrium
 state (from the top to the bottom) on a collision plane of the two
 merging galaxy clusters in the runs, 1:1 head on, 1:1 offset, 4:1
 head on, and 4:1 offset (from the left to the right), at a time of
 $t=0.5$~Gyr.\label{fig:result2_1T}}
\end{figure*}

\begin{figure}[htbp]
\begin{center}
\FigureFile(80mm,40mm){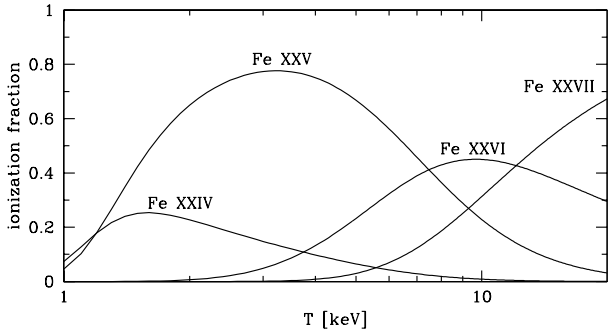}
\end{center}
\caption{Ionization fractions of Fe in the ionization equilibrium
state as a function of temperature.\label{fig:ionfrac}}
\end{figure}

As shown in the top panels of figure~\ref{fig:result2_1T}, most of ICM
except the regions behind the shocks is in the temperature range of
$5\lesssim T\lesssim 15~{\rm keV}$, and in this temperature range,
Fe\,\textsc{xxv} and Fe\,\textsc{xxvi} are dominant among all the
ionization states of iron, if ICM is in the ionization equilibrium state
(see figure~\ref{fig:ionfrac}). Therefore, in the rest of this work,
we mainly focus on the ionization fractions of Fe\,\textsc{xxv} and
Fe\,\textsc{xxvi} to address the degree of the deviation from the
ionization equilibrium in ICM.  In addition, we introduce the
``ionization temperature'', $T_{\rm ion}$, the temperature at which given
ionization fractions of Fe\,\textsc{xxv} and Fe\,\textsc{xxvi} are
realized in the ionization equilibrium state to characterize the 
ionization state of ICM. Note that $T_{\rm ion}$ monotonically
increases with the ratio of abundances of Fe\,\textsc{xxvi} relative to
Fe\,\textsc{xxv}, and differs from $\bar{T}$ if these ions are not in
the ionization equilibrium state.  Thus, the difference between
$T_{\rm ion}$ and $\bar{T}$ can be used as a measure of the deviation
from the ionization equilibrium state.  Considering the situation
where ICM is shock-heated from a few keV to higher temperature,
$T_{\rm ion}$ is lower than $\bar{T}$ just after the shock heating
since the ionization from Fe\,\textsc{xxv} to Fe\,\textsc{xxvi} is not
quick enough to catch up with the ionization equilibrium.  After that,
$T_{\rm ion}$ asymptotically approaches to $\bar{T}$ as the ionization
state gets close to the ionization equilibrium.

The middle and bottom panels of figure~\ref{fig:result2_1T} show the
maps of the ratio of the ionization temperature relative to the mean
temperature of ICM, $T_{\rm ion}/\bar{T}$, and the ratio of the fractions of
Fe\,\textsc{xxv} relative to that in the ionization equilibrium state,
$f({\rm Fe}\,\textsc{xxv})/f_{\rm eq}({\rm Fe}\,\textsc{xxv})$, at the
late stage ($t=0.5$~Gyr), respectively.  Note that the fraction in the
ionization equilibrium, $f_{\rm eq}({\rm Fe}\,\textsc{xxv})$, is
calculated using the mean temperature, $\bar{T}$.
It is clearly seen that we have a significant difference
between $T_{\rm ion}$ and $\bar{T}$ around the shock layers depicted
in figure~\ref{fig:result1_1T}, indicating that there exists a
significant deviation from the ionization equilibrium state.  The
ratio of the ionization fractions, $f({\rm Fe}\,\textsc{xxv})/f_{\rm
eq}({\rm Fe}\,\textsc{xxv})$, is above unity and $\sim 1.5$ at the
shocks in the outskirts, and $\sim 5 - 7$ and $\sim 4$ 
at the shocks in front of the ICM cores in the runs 
with 1:1 and 4:1 mass ratios, respectively.

The over-population of Fe\,\textsc{xxv} fraction at the shocks in the
outskirts and in front of the ICM cores described above can be understood as
follows. In these regions, the shock heating increases the ICM
temperature from $2 - 3$~keV to the higher temperature, so that the
Fe\,\textsc{xxv} fraction in the post-shock regions declines with time
toward the equilibrium fraction because the Fe\,\textsc{xxv} fraction
is maximum at $\sim 3$~keV in the ionization equilibrium state (see
figure~\ref{fig:ionfrac}). However, the ionization of Fe\,\textsc{xxv} 
toward higher--ionized states is not quick enough to catch up with 
the ionization equilibrium state.  Therefore, the Fe\,\textsc{xxv} fraction
is higher than that in the ionization equilibrium state.

\subsection{Two-Temperature Runs}\label{subsec2}

In this subsection, we show the results for the two-temperature runs,
in which the assumption that electrons and ions share a single
temperature is relaxed, in order to examine the effect of the
two-temperature structure on the ionization state of ICM.  The thermal
relaxation between electrons and ions is solved using equation
(\ref{eq:eq2T}). The simulations are performed for the same set of
the initial conditions as the single-temperature runs.

\begin{figure*}[htbp]
\begin{center}
\FigureFile(160mm,80mm){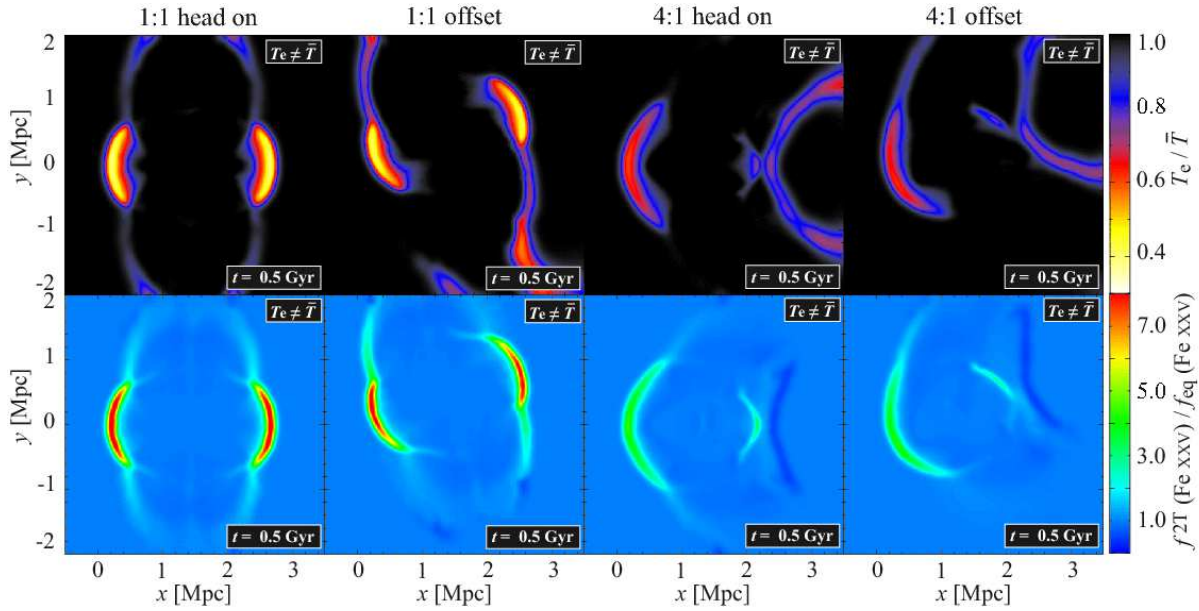}
\end{center} 
\caption{The ratio of the electron temperature relative to the mean
temperature of ICM, $T_{\rm e}/\bar{T}$, (upper) and the ratio of the
ionization fractions of Fe\,\textsc{xxv} relative to that in the
ionization equilibrium state (lower) on a collision plane of the two
merging galaxy clusters in the runs 1:1 head on, 1:1 offset, 4:1
head on, and 4:1 offset (from the left to the right), at the late
stage of the mergers ($t=0.5$~Gyr) in the two-temperature
runs.\label{fig:result1_2T}}
\end{figure*}

The upper panels of figure~\ref{fig:result1_2T} depict the maps of the
ratio of the electron temperature relative to the mean temperature of
ICM, $T_{\rm e}/\bar{T}$, on the collision plane at a time of
$t=0.5~{\rm Gyr}$. It can be seen that the electron temperature is
typically $\sim 10$--$20$~\% lower than the mean temperature at the
shocks in the outskirts, and $\sim 50$~\% and $\sim 30$~\% lower at
the shocks in front of the ICM cores in the runs with a mass ratio of 1:1
and 4:1, respectively. In the runs with 1:1 mass ratio, we have a
higher Mach number and a larger jump of the ion temperature at the
shock layers than the runs with 4:1 mass ratio, and accordingly the
larger temperature difference between electrons and ions.

Since the ionization and recombination rates in
equation~(\ref{eq:noneq}) depend on the electron temperature, the
ionization states at the shocks in the two-temperature runs are
altered from the results in the single-temperature runs in which the
ionization and recombination rates are computed using the mean
temperature of ICM, $\bar{T}$. The lower panels of
figure~\ref{fig:result1_2T} show the ratio of Fe\,\textsc{xxv}
fractions $f^{\rm 2T}({\rm Fe}\,\textsc{xxv})$ in the two-temperature
runs relative to $f_{\rm eq}({\rm Fe}\,\textsc{xxv})$
computed using $\bar{T}$. Comparing the bottom
panels of figure~\ref{fig:result2_1T} and \ref{fig:result1_2T}, we can
see that the ratios, $f^{\rm 2T}({\rm Fe}\,\textsc{xxv})/f_{\rm
eq}({\rm Fe}\,\textsc{xxv})$, are larger than 
$f({\rm Fe}\,\textsc{xxv})/f_{\rm eq}({\rm Fe}\,\textsc{xxv})$ in the
single-temperature runs at the shocks in front of the ICM cores. This is
because a delay of the heating of electrons at the shocks leaves
Fe\,\textsc{xxv} fraction higher than that in the single-temperature
runs.  In other words, the electron temperature in the two-temperature
runs (10--13~keV) is lower than that in the single-temperature
runs (15--25~keV), and thus the ionization rate of
Fe\,\textsc{xxv} to the higher ionization levels is significantly
smaller.

\subsection{Line Intensity Ratio}\label{subsec3}

Here, let us consider observational imprints of the deviation from the
ionization equilibrium and the difference in temperature between
electrons and ions in the X-ray spectroscopic observations.  X-ray
spectra of the simulated galaxy clusters are computed with the
simulated ionization state and the hydrodynamical properties of ICM by
utilizing the SPEX ver 1.10 software. In order to characterize the
signatures of the non-equilibrium ionization state, we focus on the
intensity of Fe K$\alpha$ line emissions which is usually adopted as a
diagnostic feature of the ICM metallicity in X-ray observations, and
introduce a ratio of the X-ray intensity between the two energy bands
defined as
\begin{equation}
R=\frac{I(6.6-6.7~{\rm keV})}{I(6.9-7.0~{\rm keV})},
\end{equation}
where $I(6.6-6.7~{\rm keV})$ and $I(6.9-7.0~{\rm keV})$ are mainly
contributed by the intensities of ${\rm K}\alpha$ emissions from
Fe\,\textsc{xxiv-xxv} and Fe\,\textsc{xxvi}, respectively. Since the
contribution of Fe\,\textsc{xxiv} ${\rm K}\alpha$ emission to
$I(6.6-6.7~{\rm keV})$ is minor when $f({\rm
Fe}\,\textsc{xxiv})\lesssim0.2$, $R$ roughly reflects the
ratio of K$\alpha$ line emission intensity between Fe\,\textsc{xxv}
and Fe\,\textsc{xxvi}.  Then, we introduce a ratio, $R/R_{\rm eq}$, as
a plausible observational tracer of the deviation from the ionization
equilibrium, where $R_{\rm eq}$ is the intensity ratio under the
assumption of the ionization equilibrium and thermal equipartition
between electrons and ions. Note that $R/R_{\rm eq}$ is independent of
the local ICM metallicity.

\begin{figure*}[htbp]
\begin{center}
\FigureFile(160mm,80mm){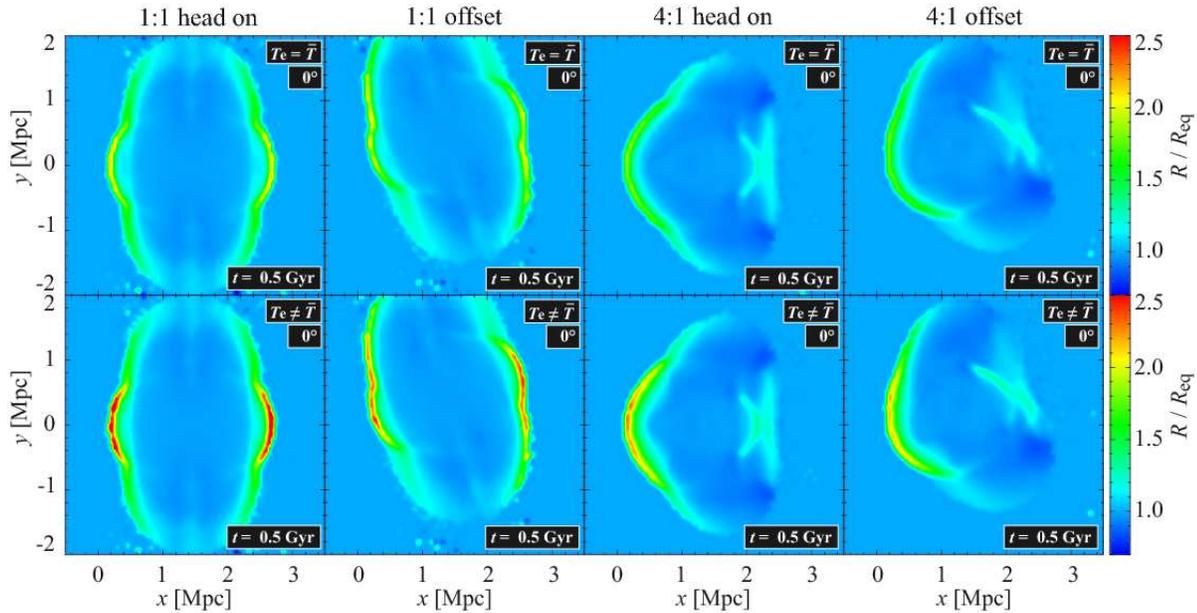}
\end{center}
\caption{The ratio, $R/R_{\rm eq}$ for the runs 1:1 head on, 1:1
offset, 4:1 head on, and 4:1 offset (from the left to the right), at
$t=0.5~{\rm Gyr}$ in the single- (upper) and two-temperature (lower)
runs.
\label{fig:intensity_ratio}}
\end{figure*}

Figure~\ref{fig:intensity_ratio} shows the maps of $R/R_{\rm eq}$ of
the simulated clusters at a time of $t=0.5~{\rm Gyr}$ in the single-
and two-temperature runs viewed along the line of sight perpendicular
to the collision plane. It is clearly seen that $R/R_{\rm eq}$
significantly exceeds unity at the shocks in the outskirts and
in front of the ICM cores. This result is consistent with the fact that
Fe\,\textsc{xxv} is over-populated compared with the ionization
equilibrium state at these shocks layers. The deviation of $R/R_{\rm
 eq}$ from unity in the two-temperature runs is larger than that in
the single-temperature runs in accordance with the fact that the
departure from the ionization equilibrium state in the two-temperature
runs is larger than that in the single-temperature runs. Furthermore,
since the difference in temperature between electrons and ions is
larger at the shocks in front of the ICM cores than at the shocks in
the outskirts, we have significant difference in $R/R_{\rm eq}$
between the single- and two-temperature runs primarily at the shocks
in front of the ICM cores.

\begin{figure*}[htbp]
\begin{center}
\FigureFile(160mm,80mm){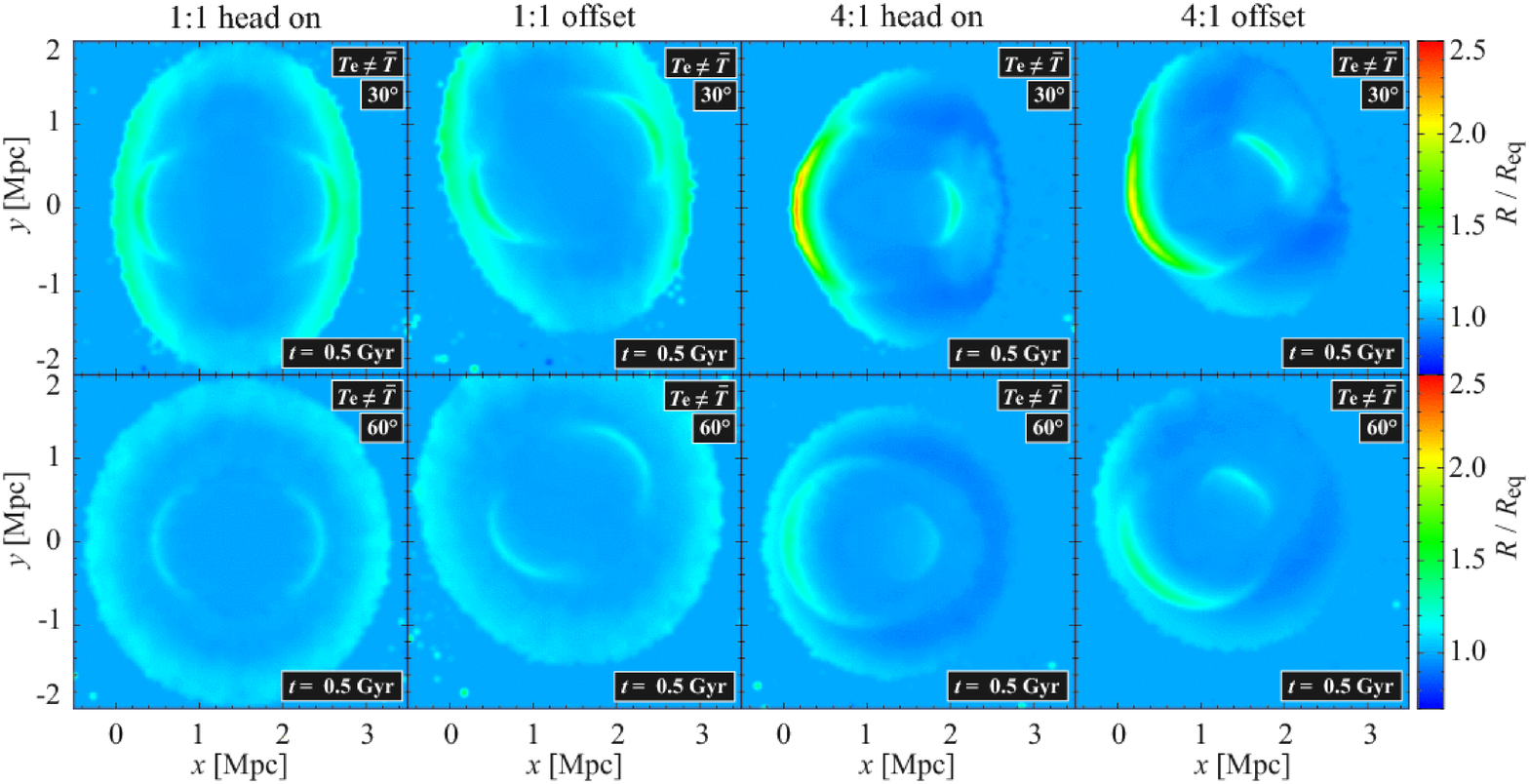}
\end{center}
\caption{Same as figure~\ref{fig:intensity_ratio} but with viewing
angles of 30 degrees (upper) and 60 degrees (lower) in the
two-temperature runs.
\label{fig:intensity_ratio_angle}}
\end{figure*}

Maps of $R/R_{\rm eq}$ in the two-temperature runs viewed from
different viewing angles (30 and 60 degrees) are shown in
figure~\ref{fig:intensity_ratio_angle}. Compared with the maps of
$R/R_{\rm eq}$ for a viewing angle of 0 degree in
figure~\ref{fig:intensity_ratio}, the deviation of $R/R_{\rm eq}$ from
unity is significantly smaller for a viewing angle of 30 degrees, and
almost negligible for a viewing angle of 60 degrees. This is because
the shock layers where the ionization state deviates from the
equilibrium state have geometrically thin structure and are almost
perpendicular to the collision plane. Therefore, when viewed with a
finite viewing angles, the signatures of the deviation from the
ionization equilibrium are diluted by the surrounding ICM which is
almost in the ionization equilibrium state.

\subsection{Projected Temperature and X-ray Surface Brightness}\label{subsec4}

Finally, let us consider the projected temperature and X-ray surface 
brightness in the merging galaxy clusters. Observed temperature of
ICM estimated by the X-ray spectrum fitting is generally different
from its physical temperature due to the projection effect along the
line of sight and the characteristic response of X-ray detectors.
Here, we adopt the spectroscopic-like temperature \citep{Mazzotta04}
as an estimate of the observed projected temperature of ICM.

\begin{figure*}[htbp]
\begin{center}
 \FigureFile(160mm,80mm){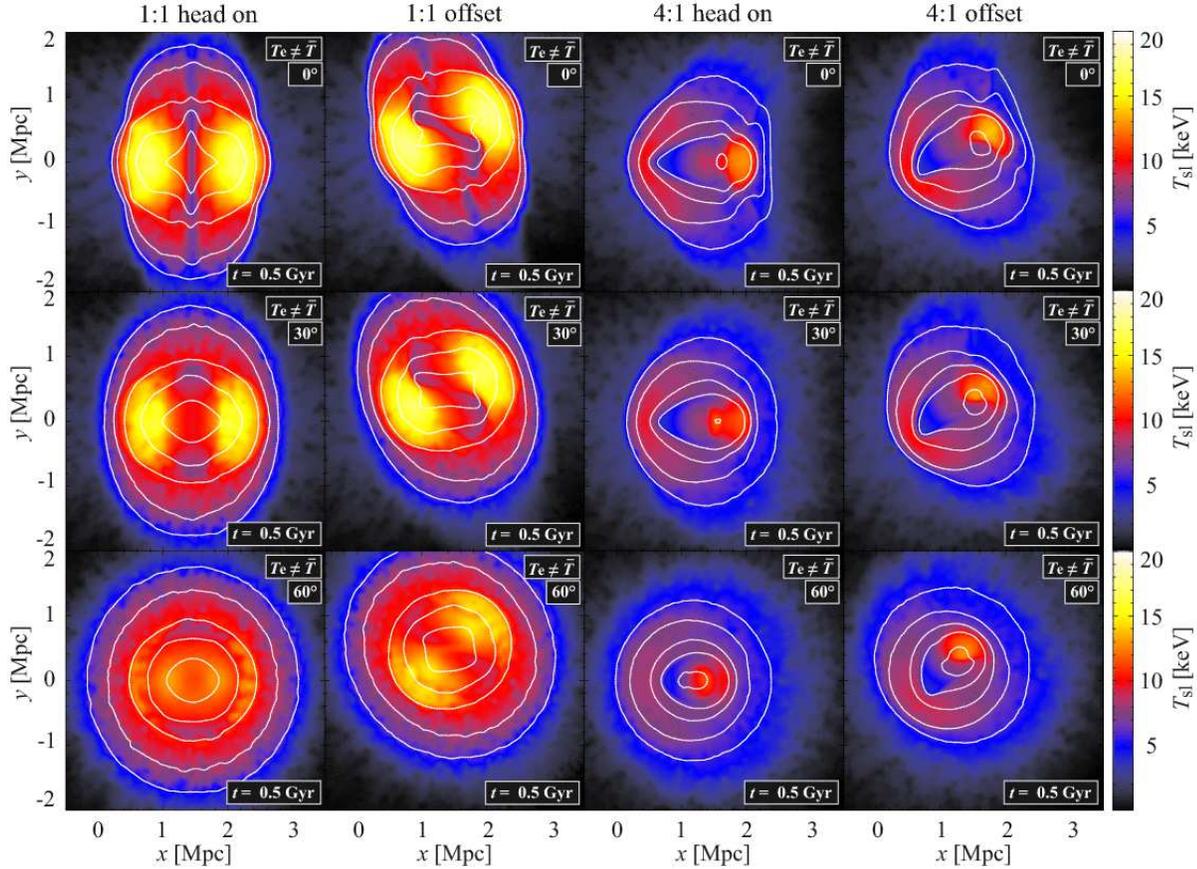}
\end{center}
\caption{The spectroscopic-like temperature map viewed with 
 a viewing angle of 0, 30, and 60 degrees (from the top to the bottom),
 respectively, for the runs 1:1 head on, 1:1 offset, 4:1 head on, 
 and 4:1 offset (from the left to the right), at $t=0.5~{\rm Gyr}$
 in the two-temperature runs. Contours shows the maps of 
 X-ray surface brightness in the 2--10 keV energy band.
 \label{fig:tmpr_brightness}}
\end{figure*}

Figure~\ref{fig:tmpr_brightness} shows the maps of the
spectroscopic-like temperature and the X-ray surface brightness in the
2--10 keV energy band viewed from different viewing angles at a
time of $t=0.5~{\rm Gyr}$.  For a viewing angle of 0 degree, jumps of
the projected temperature across the shocks are $\sim 7-13~{\rm keV}$
in front of the ICM cores and $\sim 3-7~{\rm keV}$ in the outskirts. When
viewed with a viewing angle of 30 or 60 degrees, the jumps of the
projected temperature across the shock fronts are blurred and smaller
than those in the case with a viewing angle of 0 degree, because the
shock fronts are not aligned with the line of sight.  Therefore, the
Mach numbers estimated from the jumps of the projected temperature
could be biased. Actually, the Mach numbers inferred from the apparent
jumps of the projected temperature for a viewing angles of 0, 30, and
60 degrees in the case of 1:1 head on run are 3.8, 2.2, and 1.5 while
the actual Mach number is $\sim 4$.  The jumps of projected
temperature in the single-temperature runs are almost the same as
those in the two-temperature runs, because the signatures of the 
two-temperature structure are diluted by the surrounding ICM 
which is in the thermal equipartition, when viewed with a finite viewing angle. 
Therefore, we can conclude that these biases in the Mach numbers are 
mainly due to the projection effect.

As for the X-ray surface brightness, sharp discontinuities at shock
layers are clearly visible only for a viewing angle of 0 degree. This
is again because the shock fronts are geometrically thin and the
projection effects dilute the discontinuities in the case of non-zero
viewing angles. On the other hand, contact discontinuity surfaces for
4:1 runs are visible in X-ray surface brightness when a viewing angle
is 0 degree and even 30 degrees, since the contact discontinuity
surfaces for 4:1 runs have ``bullet''-like structures and their
appearance in the X-ray surface brightness relatively insensitive to
the viewing angles.

\section{Summary and Discussion}

We carry out N-body + SPH simulations of merging galaxy clusters with
various sets of initial conditions by relaxing the assumptions of both
the ionization equilibrium and thermal equilibration between electrons
and ions in order to investigate the ionization state of ICM in the
merging galaxy clusters in detail. As for the thermal equipartition
between electrons and ions, we perform the single- and two-temperature
runs to bracket the plausible theoretical uncertainties regarding the
thermal relaxation processes between electrons and ions.

In our simulations, we find that the shocks induced by the merging
events can be classified into two types regardless of the initial
conditions. The first ones are formed at the early stage of the
merging processes in the outskirts of the clusters where the sound
velocity is roughly equal to or lower than the initial relative
velocity of the merging galaxy clusters, and the second ones are
formed in front of the ICM cores in the middle and late stages. Since
the cores of ICM is accelerated by the gravitational potential of
collisionless dark matter, the Mach number of the shocks in front of
the ICM cores is higher than that in the outskirts.

It is shown that the ionization state of iron around these shocks
deviates from the ionization equilibrium state. More specifically, the
fraction of Fe\,\textsc{xxv}, the most populated ionization state of
iron in typical galaxy clusters, exceeds its ionization fraction in
the ionization equilibrium at the shock layers both in the outskirts and 
in front of the ICM cores. This excess of the Fe\,\textsc{xxv} fraction 
can be understood
by the facts that the ionization processes at the shock layers are not
quick enough to achieve the ionization equilibrium instantaneously and
that the ionization state there differs from the equilibrium state for
a while after experiencing the shock heating.  Furthermore, in the
two-temperature runs, the electron temperature at the shock layers are
remarkably lower than the mean temperature of ICM, and thus the deviation
from the ionization equilibrium state is more significant than that in
the single-temperature runs, because the ionization processes at the
shock layers are retarded due to the lower temperature of electrons.

From the observational point of view, the deviation from the
ionization equilibrium can be identified by the difference between
intensity ratios of Fe K$\alpha$ line emissions and that inferred from
the ICM temperature estimated by the X-ray continuum spectra. We find
that the ratio, $R/R_{\rm eq}$, can be a good tracer of such a
deviation from the ionization equilibrium. We also find that the
appearance of such observational imprints of the non-equilibrium
ionization state is very sensitive to the viewing angle of the merging
galaxy clusters to the observers. $R/R_{\rm eq}$ is the largest if the
merging galaxy clusters are observed along the line of sight
perpendicular to the collision plane of the two clusters, because the
shock layers are geometrically thin and are parallel to the line of
sight. On the other hand, the significance of such imprints is weaker
in the case of a viewing angle of 30 degrees, and almost negligible in
the case of 60 degrees. It is found that jumps of the
spectroscopic-like (projected) temperature and the X-ray surface
brightness across the shock fronts also depend on the viewing angle,
and that the Mach numbers estimated from the jumps of the observed
temperature could be also underestimated in the case of a finite
viewing angle.

While we concentrate only on the ionization state of Fe so far,
detections of the deviation from the ionization equilibrium of other
heavy elements could provide independent information on physical
properties of shock waves. Among heavy elements other than Fe, Si
is the most interesting from the observational point of view because
the emission lines of Si\,\textsc{xiii-xiv} at a photon energy of
$\simeq 2$~keV are relatively prominent.  Actually, at the shocks in
the outskirts the ionization fraction of Si\,\textsc{xiv} is 1.2--1.5
and 1.2--2.0 times larger than that in the ionization equilibrium
state in the single- and two-temperature runs, respectively.  Since
the ionization fraction of Si~\textsc{xiv} in the ionization
equilibrium is maximum at a temperature of $\simeq 10^7$~K, and almost
fully ionized at higher temperature, the observational signatures of
the deviation from the ionization equilibrium could be seen in the
outskirts rather than in the central regions of galaxy clusters.

With the current X-ray observational facilities, the separate
detection of emission lines of Fe~\textsc{xxv} and Fe~\textsc{xxvi},
or the measurement of the ratio $R/R_{\rm eq}$ is still difficult
because of a lack of the spectroscopic resolution of the CCDs.
Therefore, spectroscopic analyses to estimate the metallicity of ICM
with the assumption that ICM is in the ionization equilibrium state
can lead to a biased result. A good spectroscopic resolution is also
crucial to estimate the temperatures of electrons and ions separately,
because the ion temperature is only estimated from the detailed line
profiles of emission lines, while the electron temperature can be
relatively easily estimated from the X-ray continuum spectra. Thus,
the precision spectroscopy in X-ray observations is of crucial
importance in studying the non-equilibrium ionization state and the
two-temperature structure of ICM, and can be achieved by X-ray
calorimeters on board future satellites such as Astro-H and
International X-ray Observatory.

\bigskip

This work is supported in part by Grant-in-Aid for Specially Promoted
Research (16002003) from MEXT of Japan, Grant-in-Aid for Scientific
Research (S) (20224002), for Scientific Research (A) (20340041), for
Young Scientists (Start-up) (19840008) and for Challenging Exploratory
Research (21654026) from JSPS.  Numerical simulations for the present
work have been carried out under the ``Interdisciplinary Computational
Science Program'' in Center for Computational Sciences, University of
Tsukuba. TA is supported in part by Korea Science and Engineering
Foundation (R01-2007-000-20196-0).

\end{document}